\documentclass[letterpaper, 10 pt, conference]{ieeeconf}  

\IEEEoverridecommandlockouts         

\usepackage{xcolor,calc}
\usepackage[doi=false,isbn=false,defernumbers=true,style=ieee,backend=bibtex,maxbibnames=1000]{biblatex}
\bibliography{acc2018.bib}

\usepackage[pdftex]{graphicx}
\graphicspath{{figures/}}
\usepackage{amsmath} 
\usepackage{amssymb}  
\usepackage{color}
\usepackage{enumerate}
\usepackage{multirow}
\usepackage{mathtools}
\usepackage{booktabs}
\usepackage{epstopdf}
\usepackage{tabularx}

\usepackage[normalem]{ulem}

\usepackage{algorithm,algcompatible,amsmath}

\usepackage{hyperref}
\usepackage{url}
\usepackage{bm}
\usepackage{caption}

\usepackage{booktabs}

\hypersetup{colorlinks,citecolor=black,filecolor=black,linkcolor=black,urlcolor=black}

\newtheorem{remark}{Remark}
\newtheorem{proposition}{Proposition}

\newcommand{\R}{\mathbb{R}}

\title{\LARGE \bf Adaptive MPC with Chance Constraints for FIR Systems}

\author{Monimoy Bujarbaruah, Xiaojing Zhang, Francesco Borrelli  
\thanks{M. Bujarbaruah, X. Zhang and F. Borrelli are with the Model Predictive Control Laboratory at the University of California, Berkeley, USA.
E-mails: \tt\scriptsize{ \{monimoy\_bujarbaruah, xiaojing.zhang, fborrelli\}@berkeley.edu.}} %
}

\begin{document}
\maketitle
\thispagestyle{empty}
\pagestyle{empty}

\begin{abstract}
This paper proposes an adaptive stochastic Model Predictive Control (MPC) strategy for stable linear time invariant systems in the presence of bounded disturbances. We consider multi-input multi-output systems that can be expressed by a finite impulse response model, whose parameters we estimate using a linear 
Recursive Least Squares algorithm. Building on the work of \cite{tanaskovic2013adaptive, tanaskovic2014adaptive}, our approach 
 is able to handle hard input constraints and probabilistic output constraints. By using tools from distributionally robust optimization, we formulate our MPC design task as a convex optimization problem that can be solved using existing tools. Furthermore, we  show that our adaptive stochastic MPC algorithm is persistently feasible. 
 The efficacy of the developed algorithm is demonstrated in a numerical example and the results are compared with the adaptive robust MPC algorithm of \cite{tanaskovic2014adaptive}.
\end{abstract}


\section{Introduction}
Model Predictive Control (MPC) has recently established itself as a promising tool for dealing with constrained, and possibly uncertain, systems 
\cite{mayne2000constrained, borrelli2017predictive}. 
Challenges in MPC design include presence of disturbances and/or unknown model parameters. Disturbances can be handled by means of robust or chance constraints, and such methods are generally well understood 
\cite{kothare1996robust, schwarm1999chance,Goulart2006, limon2010robust, zhang:margellos:goulart:lygeros:13, zhang2017robust, rosoliaZhangBorrelli_AR2018}. 
In this paper, we are looking into methods for addressing the second challenge posed by model uncertainties. 


If the actual model of a system is unknown, adaptive control strategies have been applied for meeting control objectives and ensuring system stability. While adaptive control for unconstrained systems is generally well-understood \cite{ioannou1996robust, sastry2011adaptive}, studies of adaptive control for systems subjected to input, state and output constraints is limited. Indeed, adaptation of system model with new available measurements can often pose intricacies in constrained optimization-based controls like MPC. One such difficulty is to  ensure the so-called recursive feasibility of the solved optimization problem \cite{borrelli2017predictive}. Due to this difficulty, adaptive control for constrained systems has mainly focused on improving performance with the adapted models, while the constraints are satisfied robustly for all possible adaptation errors and all  disturbances realizations  \cite{lorenzen2017adaptive, aswani2013provably}. 

In this paper, we build on the work of \cite{tanaskovic2013adaptive, tanaskovic2014adaptive},
and propose an \emph{Adaptive Stochastic MPC} algorithm that considers probabilistic output constraints and hard input constraints. As in \cite{tanaskovic2014adaptive}, we consider a finite impulse response model of a system that is subject to bounded disturbances with known mean and variance. The support for the set of all possible models, which we call the Feasible Parameter Set (FPS), is adapted at each time step from known error bounds on the outputs, using a set membership based approach. The main contributions of the paper can be summarized as follows:
\begin{itemize}
\item We obtain an estimate of the unknown system inside the FPS using a Recursive Least Squares (RLS) estimator. Using this estimated system, we propagate our nominal predicted outputs used in the controller objective function to improve performance. Simultaneously, we safely ensure satisfaction of the output chance constraints for the unknown true system.

\item  We show that the proposed adaptive stochastic MPC scheme is recursively feasible. That is, if the optimization problem for control synthesis has a feasible solution at time $t=0$, it continues to have a feasible solution for all subsequent $t\geq0$ under the chosen closed loop control law for the system.
\item Through numerical simulations, we demonstrate that our algorithm exhibits better performance than the algorithm presented in \cite{tanaskovic2014adaptive}.
\end{itemize}

The paper is organized as follows: in Section~\ref{sec:problem_des} we lay out the objectives and also a brief outline of the method used. In Section~\ref{sec:backg} we present the mathematical tools utilized for recursive model estimation and control synthesis. We also illustrate the cost and the constraints for our MPC controller later in this section. In Section~\ref{sec:algo} we present the proposed algorithm for the adaptive stochastic MPC controller and prove recursive feasibility. Finally, Section~\ref{sec:simul} shows the results and simulations for the developed controller along with comparisons with the existing adaptive robust MPC developed in \cite{tanaskovic2014adaptive}. We present concluding remarks and possible extensions in Section~\ref{sec:ccl_future}.

\section{Problem Description}\label{sec:problem_des}

\subsection{System Modeling and Control Objective}\label{sec:model}
We consider stable linear time-invariant systems described by a Finite Impulse Response (FIR) model of the form
\begin{align} \label{sysmodel}
y(t) = H_a \Phi(t) + w(t),
\end{align}
where the number of inputs and outputs considered is $n_u$ and $n_y$ respectively. $m$ is the length of the FIR regressor vector denoted by $\Phi(t) \in \mathcal{R}^{n_um} = [u_1(t-1), \cdots, u_1(t-m),\cdots,u_{n_u}(t-1),\cdots,u_{n_u}(t-m)]^\top$. $u_i(t)$ denotes the $i^{th}$ input at time $t$ and $y(t)\in\mathcal R^{n_y}$ is the measured output. $H_a \in \mathcal{R}^{n_y \times n_um}$ is a matrix comprising of the impulse response coefficients that relate inputs to the outputs of the system. The disturbance vector $w(t) \in \mathcal{R}^{n_y}$ is assumed to be a zero-mean random variable whose variance is known. 
The disturbance is component-wise bounded as
\begin{align}\label{disturbance_bound}
    \vert w_j(t)\vert \leq \bar w_j, \forall j = 1,2,\cdots n_y,
\end{align}
where the $\bar w_j$ are assumed known.
We wish to control the output $y(t)$ while satisfying the following input and output constraints
\begin{subequations} \label{constraints}
\begin{align}
C u(t) & \leq g, \quad t=0,1,\ldots \label{contr_con}\\
\mathcal{P}\{E y(t) \leq p\} & \geq 1-\epsilon, \quad  t=0,1,\ldots, \label{eq:cc1} 
\end{align}
\end{subequations}
where $\epsilon\in(0,1)$ is the maximum allowed probability of output constraint violation. For simplicity, we consider single linear output chance constraints. Therefore $E$ is a row vector and $p\in\R$ is a scalar. Notice that joint (linear) chance constraints can be reformulated into a set of individual (linear) chance constraints using Bonferroni's inequality, at the cost of introducing conservatism \cite{farina2016stochastic}, and can be addressed by our proposed framework.

\remark 
The FIR systems considered in this paper are suited for asymptotically stable linear systems with fast rate of impulse response decay, which is directly related to $m$, the regressor length. A small regressor length  results in fast and efficient computations. Moreover, FIR models are also valid for nonlinear systems, when they are pre-stabilized and locally linearized near an equilibrium point~\cite{tanaskovic2014adaptive}. 

\subsection{Method Outline}
We assume in this paper that the system matrix $H_a$ in (\ref{sysmodel}) is unknown. Our proposed method uses the following steps:
\begin{enumerate}
    \item At time step $t$, obtain measurement $y(t)$ and estimate $H_a$ based on past applied control inputs and measured outputs. Update the Feasible Parameter Set using known disturbance bounds (\ref{disturbance_bound}).
    \item Using the above estimate and Stochastic MPC control, compute the input sequence that satisfies constraints (\ref{constraints}) and minimizes the objective function. 
    \item  Apply the first computed control input and continue to step 1).
\end{enumerate}

In the following section, we discuss steps 1) and 2).

\section{Preparatory Material}\label{sec:backg}
We approximate system \eqref{sysmodel} with the form:
\begin{equation}\label{approxModel}
y(t)  = H(t) \Phi(t) + w(t),
\end{equation}
where model $H(t) \in \mathcal{R}^{n_y \times n_um}$ is a random variable, whose support, and first and second moments we estimate from the output measurements. The support for the set of all possible models $H(t)$, which we call the Feasible Parameter Set (FPS), also contains the true model $H_a$. In the following, we discuss how the FPS is calculated. This set is adapted with each new measurement. Based on initial known statistics, we also extract an estimate of $H_a$ in the FPS at each time step in the form of the mean of the conditional distribution of $H(t)$ given measurement $y(t)$. This value is used in the control design to improve performance. 

\subsection{Model Estimation}\label{model_est}
We obtain an estimate of the true (unknown) system $H_a$, which is characterized by a support (FPS) $\mathcal F(t) \in \mathcal{R}^{n_y \times n_um}$, mean $\mu_a(t)$ and variance $\sigma^2_a(t)$. These parameters are updated at each time step as described next.

\subsubsection*{Feasible Parameter Set (FPS) Update}\label{sec:FPSU}
Following \cite{tanaskovic2014adaptive}, a set-membership identification method is used for updating the FPS $\mathcal{F}(t)$. The initialization of $\mathcal F(0)$ is done considering the fact that the true system (\ref{sysmodel}) is stable. As detailed in \cite{tanaskovic2014adaptive}, a possible approach to defining the set $\mathcal{F}(0)$ is to impose a maximum magnitude and an exponential decay rate on the FIR coefficients. As new measurements are available at each time step, we update the FPS as given by:
\begin{multline}\label{support}
    \mathcal{F}(t)= \mathcal{F}(t-1) \cap \{H(t): H(t) \Phi(t)  \leq y (t) + \bar w\}\\
\cap \{H(t): -H(t) \Phi(t) \leq -y(t) + \bar w \},
\end{multline}
where $\bar w = [\bar w_1,\ldots,\bar w_{n_y}]^\top$ is the bound of the additive disturbance given by (\ref{disturbance_bound}). A problem of this recursive update is that the number of faces of $\mathcal{F}(t)$ can become arbitrarily large, as it grows linearly with time. Hence the memory needed for storing data can become impractical.  In order to bound the computational complexity, an alternative algorithm to compute (\ref{support}) is presented in \cite{tanaskovic2014adaptive}, which is not detailed in this paper. 



\subsubsection*{Estimating $H_a$}\label{sec: kf}

For simplicity, let us rewrite \eqref{approxModel} as
\begin{align*}
y(t) & = \bm{\Phi}(t)\mathbf{H}(t) + w(t),
\end{align*}
where $\bm{\Phi}(t) \in \mathcal{R}^{n_y \times n_yn_um}$ and $\mathbf{H}(t) \in \mathcal{R}^{n_yn_um \times 1}$ are reported in the Appendix. Furthermore, let $\sigma^2_w$ be the variance of the disturbance $w(t)$ which we assume is time invariant. Call the initial mean and variance estimates for true system $\mu_{\mathbf{a}}(0)$ and $\sigma^2_{\mathbf{a}}(0)$ respectively. Now, the conditional mean and variance estimates, given measurements up to $y(t)$, can be obtained with the standard Recursive Least Squares set of equations \cite[Sec.~(3.1)]{anderson1979}.
\subsubsection*{Projecting the Estimate}\label{project}
In general, the mean $\mu_{\mathbf{a}}(t)$ will not be in the set $\mathcal F(t)$. One way of obtaining a constrained estimate of $H_a$ in $\mathcal F(t)$ is to project the mean. As shown in \cite{simon2010kalman}, this can be achieved by solving the following optimization problem
\begin{align}\label{eq:proj}
    \mu_{\mathbf{a}}(t)= \arg\min_{X\in\mathcal F(t)} (X-\mu_{\mathbf{a}}(t))^\top M (X-\mu_{\mathbf{a}}(t)),
\end{align}
where $M>0$ is any chosen weighing matrix for the minimization. In this paper, we use $M=(\sigma^2_{\mathbf{a}}(t))^{-1}$, which results in the minimum variance filter \cite{simon2010kalman}. The mean in matrix form, that is, $\mu_a(t) \in \mathcal{R}^{n_y \times n_um}$ is obtained by reorganizing $\mu_{\mathbf{a}}(t) \in \mathcal{R}^{n_yn_um \times 1}$ into $n_um$ columns. This is used as the best estimate of the true system $H_a$, in the minimum mean squared error sense. Variance $\sigma^2_{\mathbf{a}}(t)$ keeps track of the error in estimation \cite[Sec.~3.1]{anderson1979}. 

\subsection{Control Synthesis}\label{control_synth}
\subsubsection*{Prediction Model}\label{sec:prob}
In this section, we show how the information obtained from Section~\ref{model_est} can be used in an MPC controller. Let $N>m$ be the prediction horizon for the controller. We denote the predicted outputs at time $t$ by $y(k|t) = H(t) \Phi(k|t) + w(k)$, for some $H(t)\in \mathcal{F}(t)$. $\Phi(k|t)$ will be denoted as \emph{future regressor vector}, for $k\in[t+1,t+N]$, and is computed as:
\begin{equation}\label{phi_hor}
    \Phi(k|t)= W\Phi(k-1|t) + Z u(k-1|t), \forall k\in [t+1, t+N]
\end{equation}
where, the matrices $W$ and $Z$  are reported in the Appendix (also in  \cite{tanaskovic2014adaptive}). These matrices essentially append each new predicted input in the regressor to obtain successive predicted regressor vectors of length $m$ at each step inside the horizon.  

\subsubsection*{Reformulation of Chance Constraints}
Our goal is to satisfy output chance constraints (\ref{eq:cc1}) for the true unknown system $H_a$. While designing a predictive controller, within a prediction horizon, we enforce $\mathcal P \{ E y(k|t) \leq p \} \geq 1-\epsilon$, where $y(k|t) = H(t)\Phi(k|t) + w(k)$, for some $H(t) \in \mathcal F(t)$. Therefore, to ensure satisfaction of (\ref{eq:cc1}) with unknown true system, we must satisfy it for all $H(t) \in \mathcal{F}(t)$. Now, using the theory of distributionally robust optimization \cite{calafiore2006distributionally, zymler2013distributionally}, it turns out that we can conservatively approximate the chance constraints (\ref{eq:cc1}) as follows:
\begin{multline}\label{socp1}
\kappa_{\epsilon} \sqrt{{\bar{\Phi}}^\top(k|t)\Gamma{\bar{\Phi}}(k|t)} + \Phi^\top(k|t)\bar{E}\mathbf{H}(t) -p \leq 0,\\ \forall H(t) \in \mathcal{F}(t),
\end{multline}
where we have $k\in[t+1,t+N]$, $\kappa_{\epsilon} =\sqrt{\frac{1-\epsilon}{\epsilon}}$ and $\bar \Phi(k|t) = [\Phi^\top(k|t) \quad 1 \quad 1]^\top$. Here, $\Gamma$ is an appended covariance matrix shown in the Appendix. As $\mathcal{F}(t)$ is a convex set, (\ref{socp1}) can be written as:
\begin{equation}\label{socp}
    \kappa_{\epsilon} \sqrt{{\bar{\Phi}}^\top(k|t)\Gamma{\bar{\Phi}}(k|t)} + \Phi^\top(k|t)\bar{E}f^i(t) -p \leq 0,
\end{equation}
where $f^i(t)$ denote all the vertices of the polytopic region $\mathcal{F}(t)$. Instead of using the distributionally robust approach, randomized methods \cite{zhang2015sample, grammatico2016scenario} or methods based on stochastic tubes \cite{KOUVARITAKIS20101719} can be used to reformulate (\ref{eq:cc1}) as well. 
\subsubsection*{MPC Problem}

The estimated system  $\mu_a(t)$ is used to propagate the nominal predicted states which are utilized in the cost function. Then we solve the following optimization problem for given $Q \in \mathcal{R}^{n_y \times n_y},S \in \mathcal{R}^{n_u \times n_u}>0$: 
\begin{equation}\label{mpc_problem}
    \begin{array}{llll}
        \displaystyle\min_{U(t)} & \displaystyle \sum_{k=t}^{t+N-1}  [\hat{y}^\top(k|t)Q\hat{y}(k|t)+u^\top(k|t)Su(k|t)] \\
       & \qquad \qquad \qquad \qquad \quad \quad + \hat{y}^\top(t+N|t)Q\hat{y}(t+N|t) \\
        \text{s.t.} & \hat{y}(k+1|t) = \mu_a(t)\Phi(k+1|t), \\[0.7ex]
         & \hat{y}(t|t) = y(t) \\[0.7ex]
        & Cu(k|t) \leq g \\[0.7ex]
        & \Phi(t+N|t)  =W\Phi(t+N|t)+Z u(t+N-1|t) \\ [0.7ex]
        & \kappa_{\epsilon} \sqrt{{\bar{\Phi}}^\top(k+1|t)\Gamma{\bar{\Phi}}(k+1|t)}\\
        & \qquad \qquad \qquad \qquad \quad +\Phi^\top(k+1|t)\bar{E}f^i(t)\leq p \\ [0.7ex]
        & \forall k = t, \ldots, t+N-1 \\ [0.7ex]
        & \forall f^i(t) \in \textrm{vertex}(\mathcal{F}(t)),
    \end{array}
\end{equation}
where $U(t) = [u(t|t)^\top, u(t+1|t)^\top, \ldots, u(t+N-1|t)^\top]^\top$, and the regressor $\Phi(k|t)$ is as in (\ref{phi_hor}). Note that the objective function minimized is with respect to the estimated system. 

We have included the terminal constraint on the regressor vector as given in \cite{tanaskovic2014adaptive}:
\begin{align}\label{eq:trc}
\Phi(t+N|t)=W\Phi(t+N|t)+Z u(t+N-1|t).
\end{align}
This means the terminal regressor corresponds to a steady state, that is, last $m$ control inputs in a horizon are kept constant. Problem (\ref{mpc_problem}) is a convex optimization problem and can be solved  with existing solvers \cite{Boyd:2004:CO:993483}. 

\section{Adaptive Stochastic MPC Algorithm}\label{sec:algo}
Let 
\begin{multline*}
 U^*(t) = [u^*(t|t)^\top, u^*(t+1|t)^\top, \cdots,u^*(t+ N-1|t )^\top]^\top
\end{multline*}
be the solution of (\ref{mpc_problem}) at time $t$. In Model Predictive Control (MPC), the first  input $u^*(t|t)$ of $U^\star(t)$ is applied to the system (\ref{sysmodel}), i.e.,
\begin{equation}\label{eq:RHC}
    u(t)=u^*(t|t).
\end{equation}
At the next time step, we resolve the optimization problem (\ref{mpc_problem}) with new estimated data $\mu_a(t+1)$ and $\mathcal F(t+1)$. This yields a receding-horizon control scheme.
The resulting algorithm is summarized in Algorithm~1. 

\begin{algorithm}
    \caption{
Adaptive Stochastic MPC
    }
    \label{alg1}
        \begin{algorithmic}[1]
      \STATE Set $t=0$; initialize mean $\mu_{\mathbf{a}}(0)$, variance $\sigma^2_{\mathbf{a}}(0)$ and support $\mathcal F(0)$. 
           
            \STATE Compute $U^*(t)$ from (\ref{mpc_problem}) and apply $u(t) = u^*(t|t)$ to the system (\ref{sysmodel}).
            
            
            \STATE 
            Obtain output $y(t+1)$, and update support $\mathcal F(t+1)$ as given in (\ref{support}).
           Estimate mean and variance $\mu_{\mathbf{a}}(t+1)$ and $\sigma^2_{\mathbf{a}}(t+1)$ with the RLS estimator. Project the mean to the known support $\mathcal F(t+1)$ as in \eqref{eq:proj}. 
            
            \STATE Set $t = t+1$, and return to step~2.
        \end{algorithmic}
\end{algorithm}

\proposition Consider Algorithm~1 and the receding horizon closed loop control law (\ref{eq:RHC}) applied to system (\ref{sysmodel}) after solving optimization problem (\ref{mpc_problem}). If the optimization problem (\ref{mpc_problem}) is feasible at time $t=0$, then it is feasible at all subsequent times $t\geq0$. 


\begin{proof} After solving (\ref{mpc_problem}) and applying (\ref{eq:RHC}) in closed loop at $t$, consider an open loop control sequence at the next time step $t+1$ as:
\begin{multline} \label{feasinput}
   U(t+1) = [u^*(t+1|t)^\top,...,u^*(t+ N-1|t )^\top, \\
   u^*(t + N-1|t)^\top]^\top.
\end{multline}
It is clear that control sequence (\ref{feasinput}) is feasible at $t+1$ as it satisfies (\ref{contr_con}). Using (\ref{feasinput}) and condition (\ref{eq:trc}), we obtain:
\begin{subequations} \label{phic}
\begin{align} 
\Phi(k|t +1) & = \Phi^*(k|t), \forall k\in [t+2,t +N]\\
\Phi(t +N +1|t +1) & = \Phi^*(t +N|t),
\end{align}
\end{subequations}
To show recursive feasibility of (\ref{mpc_problem}), first we must have (\ref{socp}) satisfied with (\ref{feasinput}) at $t+1$. That is, we require $\forall k\in[t+2,t+N+1]$:
\begin{multline}\label{rfc}
    \kappa_{\epsilon} \sqrt{{\bar{\Phi}}^\top(k|t+1)\Gamma{\bar{\Phi}}(k|t+1)} \\+ \Phi^\top(k|t+1)\bar{E}f^i(t+1)-p \leq 0
\end{multline}
$\forall f^i(t+1) \in \mathcal{F}(t+1)$ vertices. Now, for the chosen input sequence (\ref{feasinput}), by using (\ref{eq:trc}) and (\ref{phic}), condition (\ref{rfc}) can be expressed as $\forall k\in[t+2,t+N]$:
\begin{equation}\label{rfc_final_socp}
    \kappa_{\epsilon} \sqrt{{\bar{\Phi}}^{*\top}(k|t)\Gamma{\bar{\Phi}}^*(k|t)} + \Phi^{*\top}(k|t)\bar{E}f^i(t+1) -p \leq 0.
\end{equation}
We can now guarantee that (\ref{rfc_final_socp}) will be satisfied at $t+1$ if the MPC problem (\ref{mpc_problem}) is feasible at $t$. This is due to the observation that feasible parameter set follows $\mathcal{F}(t)\geq \mathcal{F}(t+1)$ as new cuts (\ref{support}) are introduced at each time step. So vertices $f^i(t+1)$ at $t+1$ are convex combinations of the ones at $t$. Thus the MPC problem (\ref{mpc_problem}) is recursively feasible under closed loop control law (\ref{eq:RHC}). This completes the proof.
\end{proof}
\remark
Instead of formulating the second order cone constraints as (\ref{socp}) for all $H(t) \in \mathcal{F}(t)$, one could also formulate them using the estimate of $H_a$ given by $\mu_a(t)$ and the corresponding variance $\sigma^2_a(t)$. This leads to imposing the chance constraints (\ref{eq:cc1}) with only an estimated probability distribution function of the true system. 
Even though that potentially relaxes ``conservatism", the chance constraints might not be satisfied in practice for the true system $H_a$. 
\remark While reformulating (\ref{socp}) for satisfaction of (\ref{eq:cc1}) with an estimated Probability Distribution Function of true system is possible (Remark~1), proving recursive feasibility of MPC in this case becomes more complicated, as the estimate $\mu_a(t)$ can change arbitrarily at each time step. This is subject to current investigation. 

\section{Simulation Results}\label{sec:simul}
    We present simulation results in this section for a simple single-input single-output system. We compare results from our adaptive stochastic MPC with the ones from the adaptive robust MPC presented in \cite{tanaskovic2014adaptive}. For simulating both the algorithms, we use the parameters given in Table~\ref{table:par}. 
    
    In contrast to our algorithm that satisfies output constraints probabilistically and optimizes the cost with respect to the estimated system, the adaptive robust MPC algorithm in \cite{tanaskovic2014adaptive} only utilizes the FPS information for robust constraint satisfaction. In \cite{tanaskovic2014adaptive} all output constraints are treated as hard constraints and are imposed for all possible system models in the FPS. Also, there is no data-driven estimation done for learning the true model. Instead, the algorithm in \cite{tanaskovic2014adaptive} uses the Chebyshev center $H_{\textnormal{cc}}(t)$ of the FPS at each time step to propagate nominal open loop outputs in the cost function.
    
    We run $100$ 
    simulations with both algorithms for $100$ randomly chosen disturbance sequences. We compare costs and outputs from both to demonstrate the effect on performance of potentially tolerating output constraint violations. We also show characteristics of the RLS estimator and illustrate its positive effect on performance. 
\subsection*{Cost Comparison}  

    \begin{table}[!t]
	\renewcommand{\arraystretch}{1.3}
	\caption{Simulation Parameters}
	\label{table:par}
	\centering
	 \begin{tabular}{c c | c c}
		\hline
		\bfseries Parameter & \bfseries Value & 
		\bfseries Parameter & \bfseries Value \\
		\hline
		$m$ & $3$ & $N$ & $12$ \\
		$t$ & $[0,20]$ & $\Phi(0)$ & $[2,2,2]^\top$ \\
		$n_u$ & $1$ & $n_y$ & $1$ \\
		$\epsilon$ & $0.3$ & $\kappa_\epsilon$ & $1.5275$\\
		$\bar w$ & $1$ & $w$ & $\textit{U}\sim[-1,1]$\\
		$E$ & $1$ & $p$ & $1$ \\
		$C$ & $\textbf{diag}(1,-1)$ & $g$ & $[3,3]^\top$\\ 
		$Q$ & $\textbf{diag}(2,2)$ & $S$ & $\textbf{diag}(2,2)$\\ 
		$\mu_{\mathbf{a}}(0)$ & $[-3,5,-4]^\top$ & $\sigma^2_{\mathbf{a}}(0)$ & $\textbf{diag}(1,1,1)$\\ 
		\hline
	\end{tabular}
\end{table}

\begin{figure}[!h] 
	\centering
	\includegraphics[width=\columnwidth]{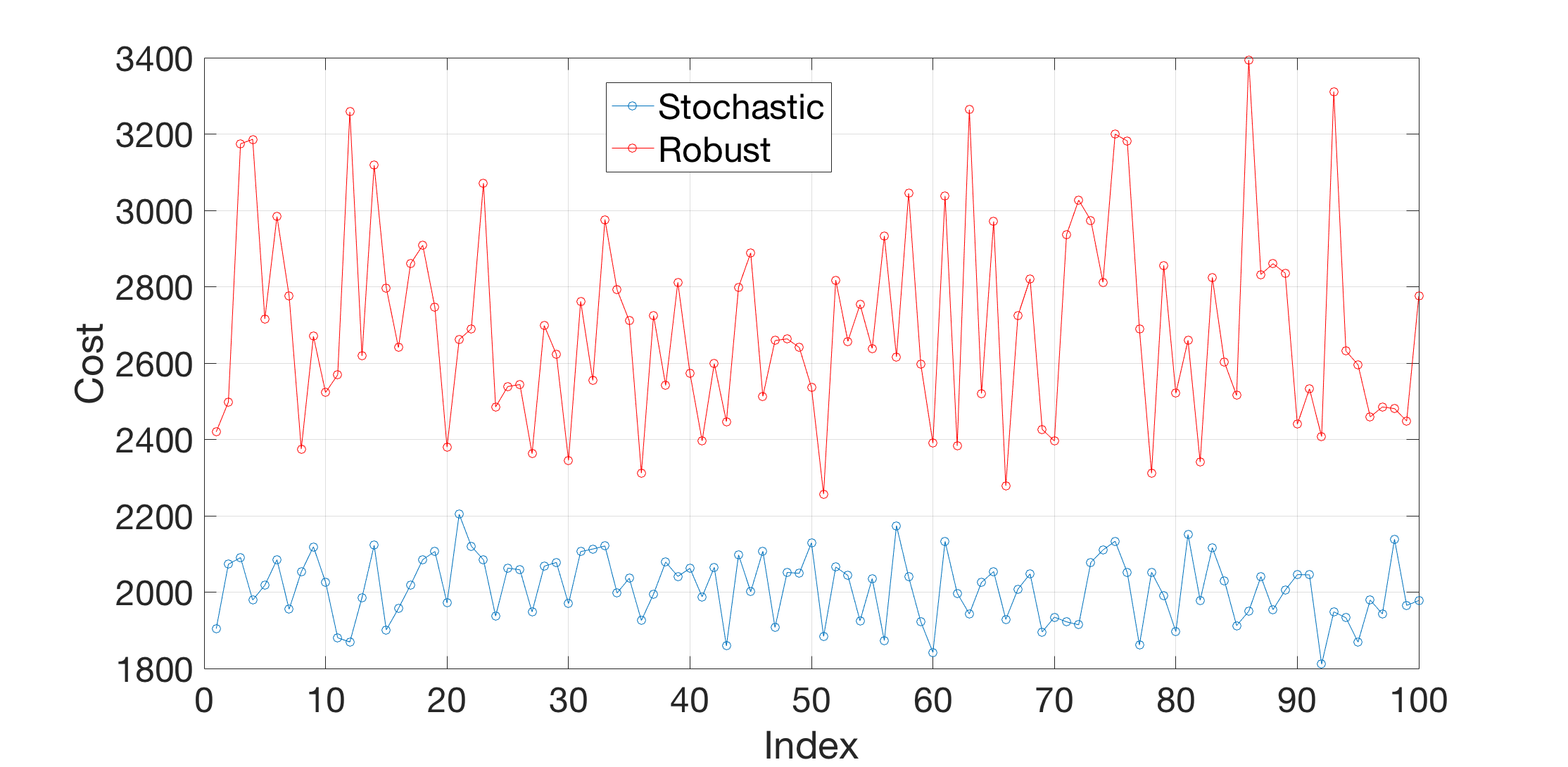}
	\caption{Cost Comparison}
	\label{fig:cost}
\end{figure}

The cost values  (measured in terms of the objective function) obtained with both algorithms are plotted in Fig.~\ref{fig:cost}. It is observed that for the simulations run with identical disturbance sequences on both systems, our adaptive stochastic MPC yields a significantly lower cost value than the adaptive robust MPC of \cite{tanaskovic2014adaptive}, underscoring better performance. 
This indicates towards two important inferences: $(i)$ allowing constraints to be violated with a small probability can be rewarding in terms of cost, and $(ii)$ our proposed algorithm yields better estimate of the true system dynamics with the RLS estimator. Moreover, unlike the Chebyshev center, the RLS estimator gives quantifiable minimum variance measure of how far our estimate is from the true system $H_a$.

\subsection*{Model Estimation}
\begin{figure}[!h] 
	\centering
	\includegraphics[width=\columnwidth]{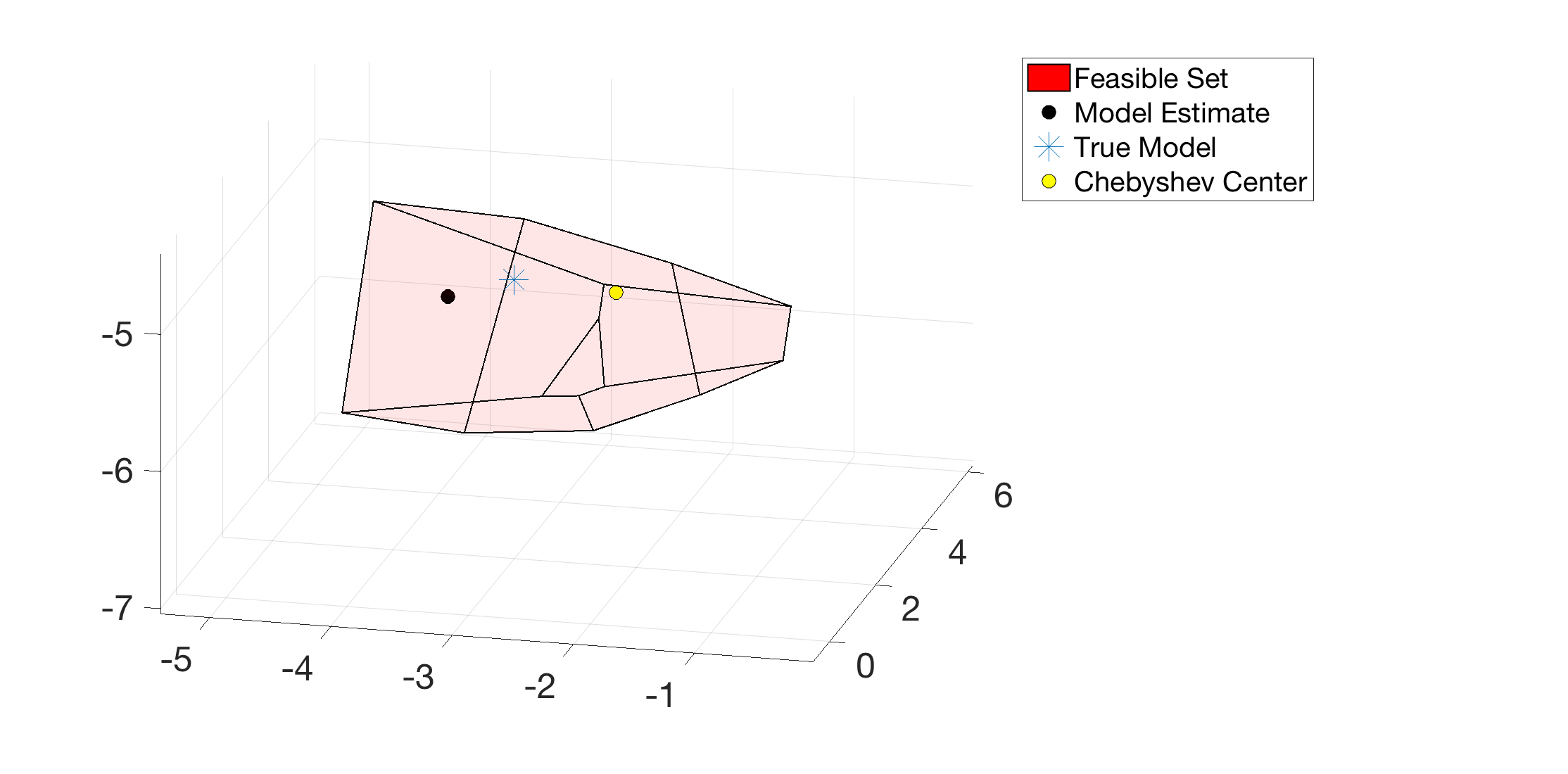}
	\caption{Final Model Properties}
	\label{fig:HF_MB}
\end{figure}

In the following, we analyze the model estimation and its convergence characteristics. True model $H_a = [-4,8,-9]^\top$ is generated here purely for the purpose of simulations, as it is not actually known. Fig.~\ref{fig:HF_MB} shows the converged (after $20$ time-steps in closed loop) asymptotic estimate $\mu_{\mathbf{a}}(\infty)$ and the Chebyshev center $H_{\textnormal{cc}}(\infty)$ of the polytope $\mathcal F(\infty)$ for one particular simulation.  

\begin{figure}[!h] 
	\centering
	\includegraphics[width=\columnwidth]{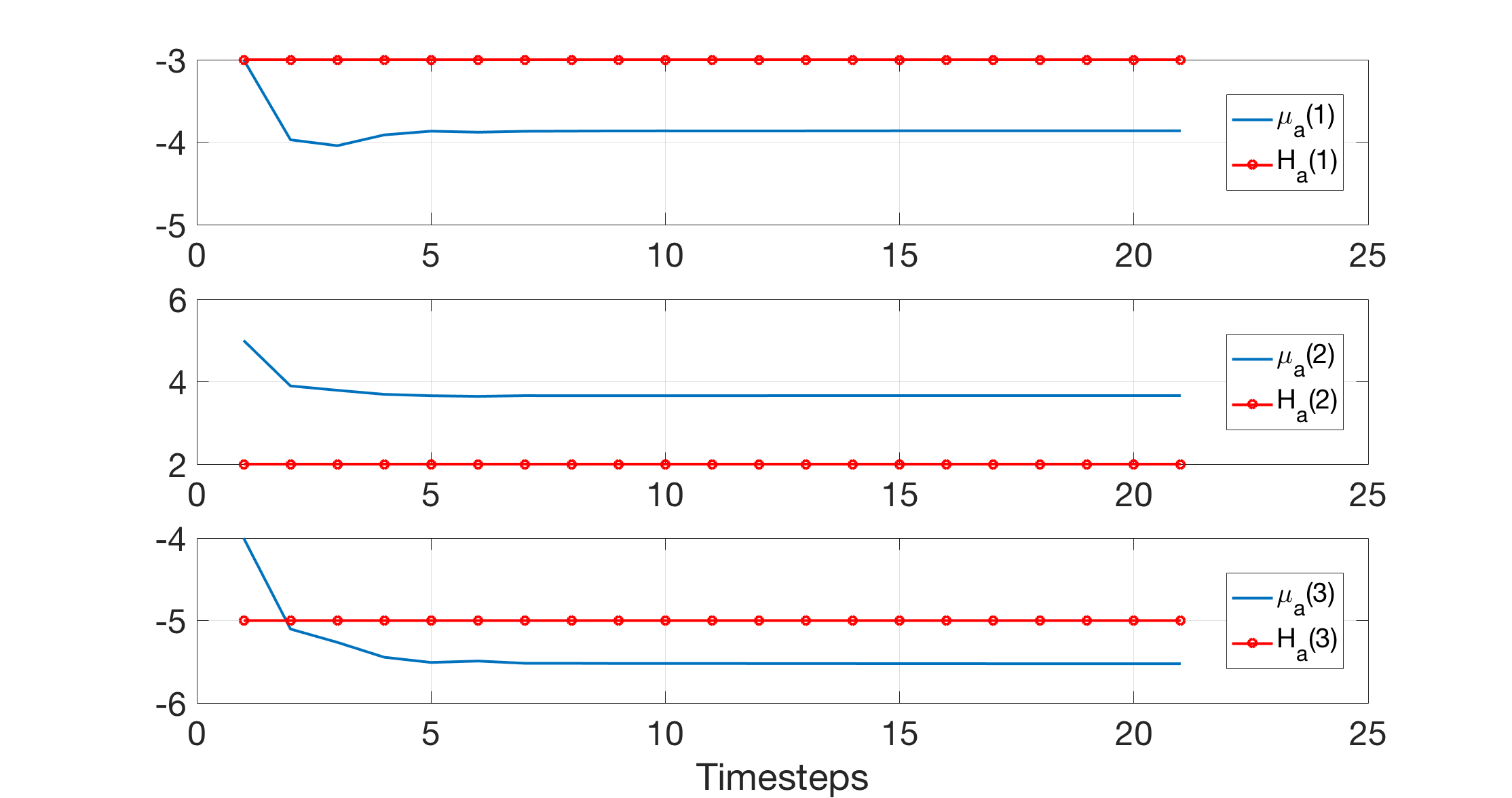}
	\caption{Model Estimate Evolution}
	\label{fig:mu_var_MB}
\end{figure}

It can be observed in the representative example from Fig.~{\ref{fig:HF_MB}} that the estimate $\mu_{\mathbf{a}}(\infty)$ approaches closer to the true model $H_a$, as compared to $H_{\textnormal{cc}}(\infty)$. This indicates that propagating the nominal predicted outputs with the RLS estimate is more accurate than with the Chebyshev center for this case. Fig.~\ref{fig:mu_var_MB} also shows the evolution of the model estimate $\mu_a(t)$ over time.

\begin{figure}[!h] 
	\centering
	\includegraphics[width=\columnwidth]{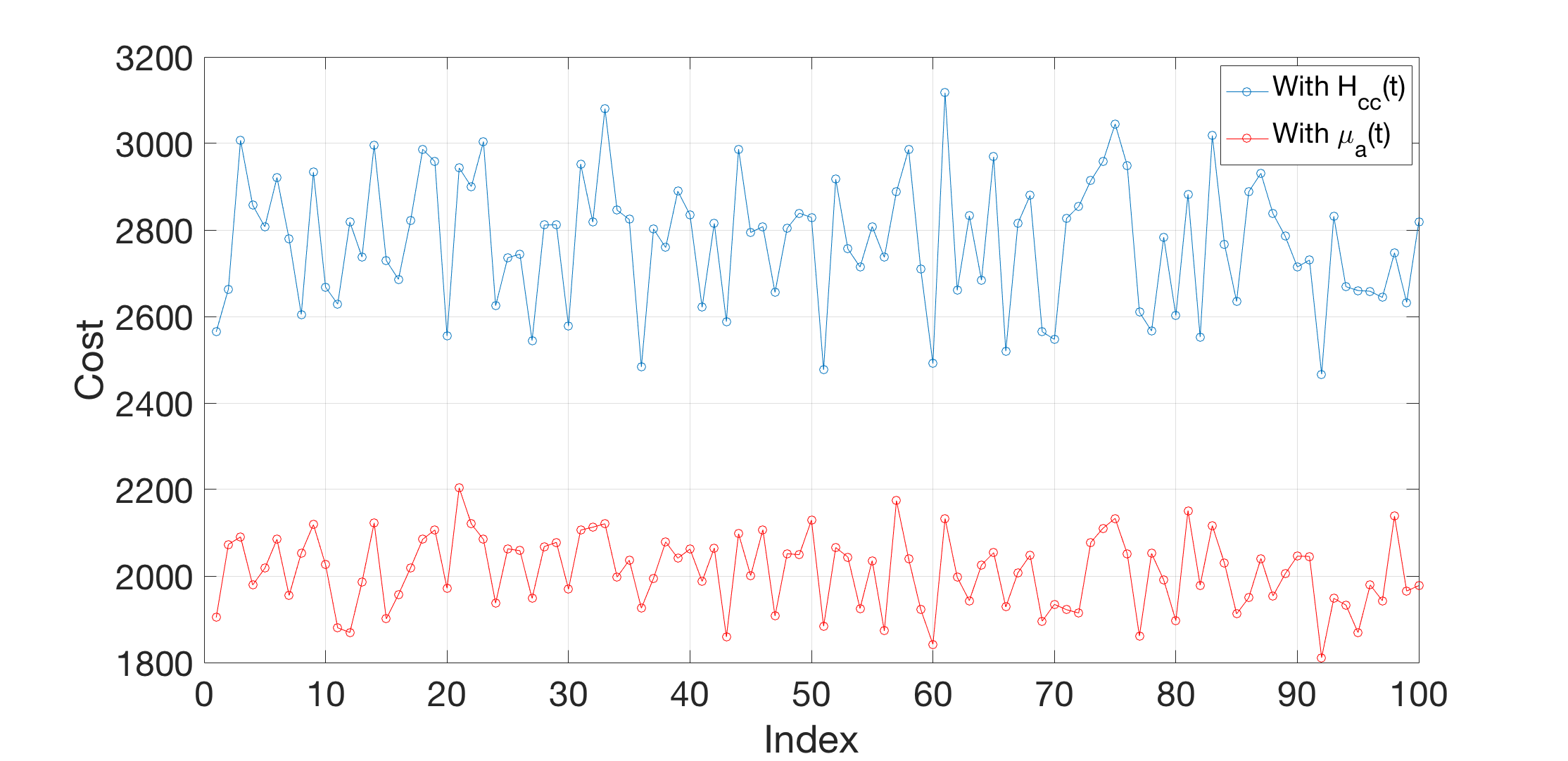}
	\caption{Performance Gain with RLS Estimator}
	\label{fig:cost_CR}
\end{figure}

The efficacy of approximating the true model $H_a$ with the RLS estimate over the Chebyshev center at each time $t$, is also highlighted in Fig.~\ref{fig:cost_CR}. Here we compare the cost of our adaptive stochastic MPC with two different model estimates: $(i)$ the RLS estimate $\mu_a(t)$ and $(ii)$ the Chebyshev center $H_{\textnormal{cc}}(t)$. We see that the performance is significantly better with the RLS estimate used as the nominal model.


\subsection*{Output Constraint Violation}
\begin{figure}[!h] 
	\centering
	\includegraphics[width=\columnwidth]{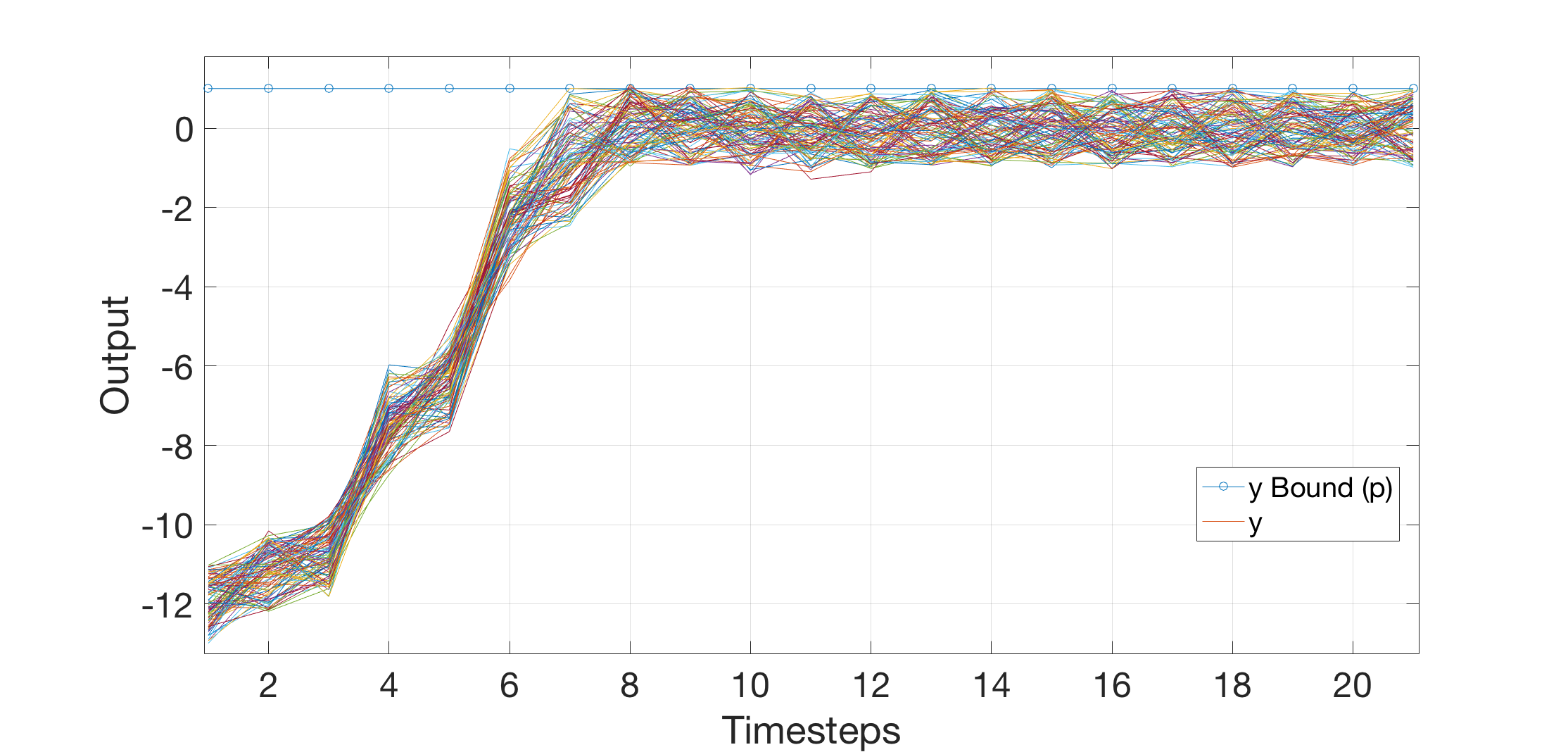}
	\caption{Output: Adaptive Stochastic MPC}
	\label{fig:y_MB}
\end{figure}

\begin{figure}[!h] 
	\centering
	\includegraphics[width=\columnwidth]{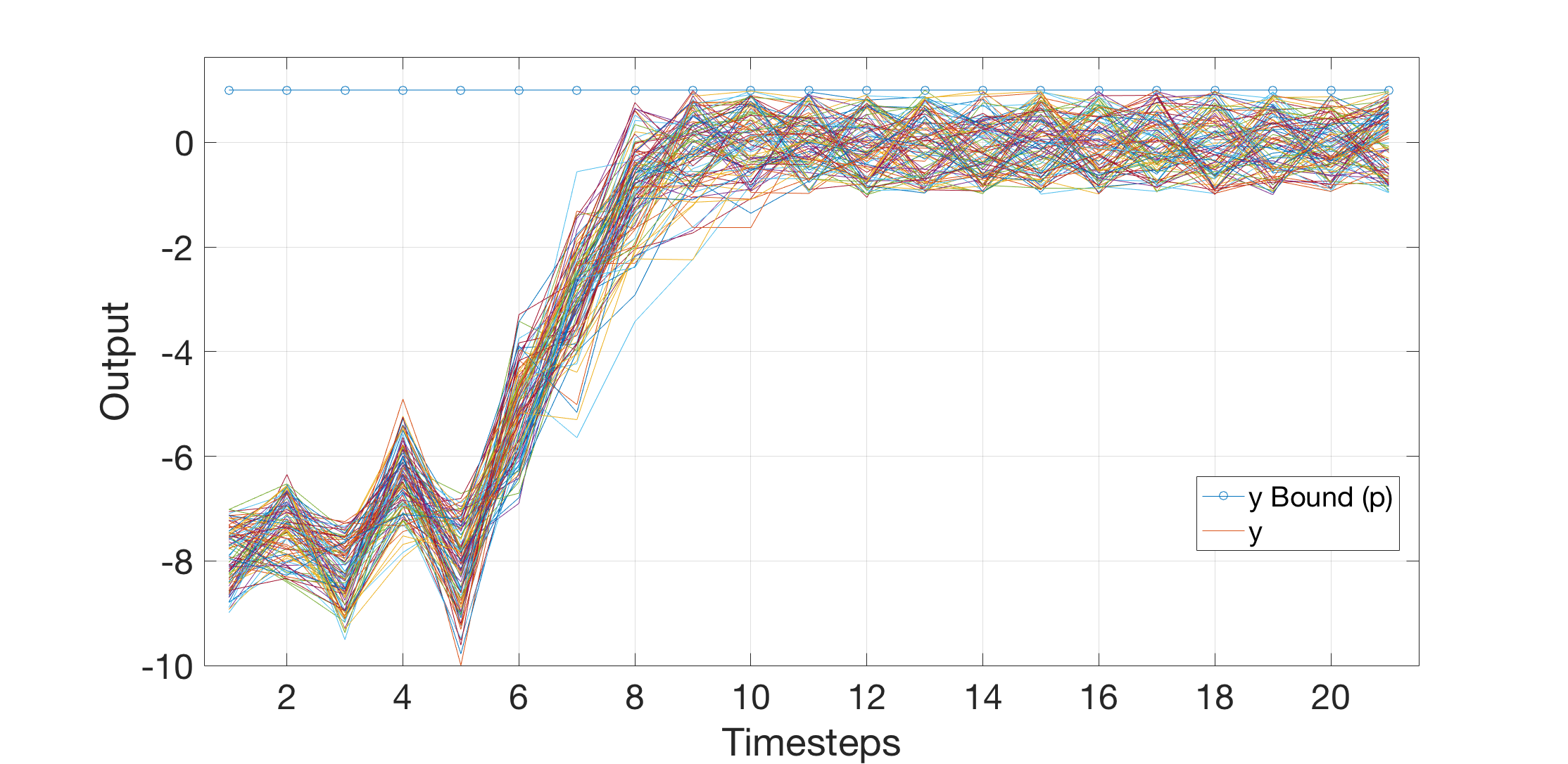}
	\caption{Output: Adaptive Robust MPC }
	\label{fig:y_MT}
\end{figure}

 Fig.~\ref{fig:y_MB} and Fig.~\ref{fig:y_MT} illustrate the comparison of closed loop output $y$ for both algorithms, under the previously chosen disturbance sequences. 
 We see that output constraints are robustly satisfied in the case of adaptive robust MPC \cite{tanaskovic2014adaptive} in Fig.~{\ref{fig:y_MT}}. Moreover, from $10\,000$ Monte Carlo simulations, we get a maximum empirical constraint violation probability of approximately $0.19$ with our adaptive stochastic MPC. Calculations show this holds true in Fig.~\ref{fig:y_MB}, where adaptive stochastic MPC respects the allowed maximum constraint violation probability of $\epsilon=0.3$. Thus, the improved performance of adaptive stochastic MPC as seen in Fig.~\ref{fig:cost}, does not come with the price of excessive constraint violations.

    \section{Conclusion and Future Work}\label{sec:ccl_future}
    We developed an adaptive stochastic MPC algorithm for stable linear time invariant systems with additive uncertainty. Our algorithm can deal with hard constraints on inputs and chance constraints on outputs. The chance constraints are enforced for all feasible models, thus ensuring that the unknown true system also satisfies them. We have guaranteed recursive feasibility of the MPC algorithm. We use the Minimum Mean Squared Error estimate of unknown system model to improve performance. 
    We also compared results of our adaptive stochastic MPC with the adaptive robust MPC algorithm by  \cite{tanaskovic2014adaptive}. For the cases simulated with identical disturbance sequences, we have observed lower cost with our algorithm while satisfying the chance constraints at all times.
    
    A future expansion of this work would extend our framework to reference tracking problems, as that can be useful for constrained tracking under potential model uncertainties as \cite{bujarbaruahlyap}. The work additionally demands designing better nonlinear estimators and quantifying confidence intervals for lesser conservatism in control design. Eventually, we wish to employ our algorithm to torque based driver-in-the-loop steering assistance systems in semi-autonomous cars  \cite{bujarbaruahtorque}.      
    
    \section*{Appendix}
    \subsection{Matrix Notations}
We define
\begin{align*}
    \mathbf{H}(t) & = [H_{1}, H_{2}, \cdots, H_{n_y}]^\top \in \mathcal{R}^{n_yn_um \times 1}\\
    \bm{\Phi}(t) & = \textbf{diag}(\Phi^\top(t),\Phi^\top(t),...,\Phi^\top(t))\in \mathcal{R}^{n_y \times n_yn_um},
\end{align*}
where $H_i(t)$ denotes the $i^{th}$ row of $H(t)$. Moreover, 
\begin{align*}
q & = \left[\begin{array}{ccccc}
0&0&\cdots &0&0\\
1&0&\cdots &0&0\\
0 &1 &\cdots &0&0\\
\vdots &\vdots &\ddots &\vdots&\vdots\\
0&0&\cdots &1&0\\
\end{array}\right] \in \mathcal{R}^{m \times m}
\end{align*}
Based on this matrix $q$ we get: 
\begin{align*}
W & =  \textbf{diag}(q,q,...,q) \in \mathcal{R}^{n_um \times n_um}
\end{align*}
\begin{align*}
    z & = [1, 0, \cdots, 0]^\top \in \mathcal{R}^m, \quad \text{which gives,}
\end{align*}
\begin{align*}
Z & =  \textbf{diag}(z,z,...,z) \in \mathcal{R}^{n_um \times n_u}
\end{align*}

\subsection{Chance Constraint to Convex Cone}
As we have stated previously, to ensure satisfaction of (\ref{eq:cc1}) with unknown true system, we must satisfy them for all $H(t) \in \mathcal{F}(t)$. Thus $H(t)$ is treated as a deterministic variable (mean $H(t)$ and variance $\mathbf{0}$) while reformulating (\ref{eq:cc1}) to equivalent convex constraints. Therefore, $\forall H(t) \in \mathcal F(t)$, we have: 
\begin{align*}
\mathcal{P}\{EH(t)\Phi(k|t) + Ew(k) \leq p\} & \geq 1-\epsilon\\
\iff \mathcal{P}\{[EH(t) \quad Ew(k)][\Phi^\top(k|t) \quad 1]^\top \leq p\} & \geq 1-\epsilon. 
\end{align*}
Let's denote, 
\begin{align*}
 a_1^\top(t) & = [EH(t) \quad Ew(k)], \implies \hat{a}_1^\top(t) & = [EH(t) \quad 0], 
\end{align*}
where $\hat{x}$ is used to denote mean of a quantity $x$. Also,
\begin{align*}
 \bar{\Phi}(k|t) & = [\Phi^\top(k|t) \quad 1 \quad 1]^\top, \quad \text{and,}
\end{align*}
\begin{align*}
 d_1(t) & = [a_1^\top(t) \quad -p]^\top, \implies \hat{d}_1(t) & = [\hat{a}_1^\top(t) \quad -p]^\top.
\end{align*}
From these we can derive the variance $\Gamma$ of $d_1(t)$ as:
\begin{align*}
     \Gamma & = \sigma^2(d_1(t))\\
     &  = \sigma^2(\begin{bmatrix}
         \bar{E}{\mathbf{H}(t)} \\
         E {w}(k)\\
         -p\\
        \end{bmatrix}), \quad \text{where }
     \bar E = \textbf{diag}(E,E,...,E)\\
& = \textbf{diag}(0, E\sigma^2_wE^\top,0)
\end{align*}
we also assume no correlation between the disturbance and the impulse response distribution in the above derivation. Now, (\ref{eq:cc1}) inside an MPC horizon can be written as convex second order cone constraints. They are given by $\forall k\in[t+1,t+N]$,
\begin{equation}\label{socp11}
\kappa_{\epsilon} \sqrt{\{{\bar{\Phi}}^\top(k|t)\Gamma{\bar{\Phi}}(k|t)}\} + \hat{d}_1^\top(t){\bar{\Phi}}(k|t) \leq 0,
\end{equation}
where $\kappa_\epsilon = \sqrt{\frac{1-\epsilon}{\epsilon}}$ for any bounded disturbance distributions $w(k)$ with known moments. After simplifications, (\ref{socp11}) can be written for all $H(t) \in \mathcal F(t)$ as:
\begin{equation*}
\kappa_{\epsilon} \sqrt{\{{\bar{\Phi}}^\top(k|t)\Gamma{\bar{\Phi}}(k|t)}\} + \Phi^\top(k|t)\bar{E}\mathbf{H}(t) -p \leq 0.
\end{equation*}

\section*{Acknowledgement}
The authors would like to thank Dr. Marko Tanaskovic for stimulating discussions.


\renewcommand{\baselinestretch}{0.930}
\renewcommand{\baselinestretch}{1}

\section*{References}
{
\printbibliography[heading=none, resetnumbers=true]
}


\end{document}